\documentclass[useAMS,usegraphicx,usenatbib]{mn2e}

\usepackage{times}

\input{epsf.tex}

\title[The hidden structure of the merger galaxy NGC 1316]
{Looking inside the nest: the hidden structure of the merger galaxy NGC 1316 (Fornax A)\thanks{Based on data collected by the New Technology Telescope -- ESO La Silla Observatory.}}
\author[Y. Beletsky et al.]
{Y. Beletsky$^{1}$\thanks{E-mail: ybialets@eso.org},
D. A. Gadotti$^{1}$,
A. Moiseev$^{2}$,
J. Alves$^{3}$,
A. Kniazev$^{4}$\\
$^{1}$European Southern Observatory (ESO), Alonso de Cordova 3107, Santiago, Chile\\
$^{2}$Special Astrophysical Observatory, Russian Academy of Sciences, Nizhnii Arkhyz,
Karachai-Cherkessian Republic, 357147 Russia\\
$^{3}$Institute of Astronomy, University of Vienna, Turkenschanzstr. 17, 1180 Vienna, Austria\\
$^{4}$South African Astronomical Observatory, PO Box 9, 7935, Cape Town, South Africa\\}

\begin{document}

\date{Accepted. Received.}

\pagerange{\pageref{firstpage}--\pageref{lastpage}} \pubyear{}

\maketitle

\label{firstpage}

\begin{abstract}
We present an analysis of the circumnuclear structure of NGC 1316 using both
near-infrared imaging and stellar kinematics. 2D decomposition of the images suggests the presence of a structure that resembles inner {\em gaseous} spiral arms, at about 5 to 15" from the center ($\approx$ 500 to 1500 pc). We also find a disc-like {\em stellar} structure with radius less than 200 pc. Analysis of previously published
SINFONI integral field kinematics data indicates a kinematically decoupled core in the same spatial scale, further evidence that indeed the nuclear stellar structure found is a kinematically cold stellar disc. We suggest that both newly-found structural components are the result of a recent accretion of a companion galaxy.
\end{abstract}

\begin{keywords}
galaxies: evolution -- galaxies: formation -- {\bf galaxies: individual: (NGC~1316=Fornax~A)} -- galaxies: interactions -- galaxies: peculiar -- galaxies: structure
\end{keywords}

\begin{figure*}
\includegraphics[width=6.5in,clip=true]{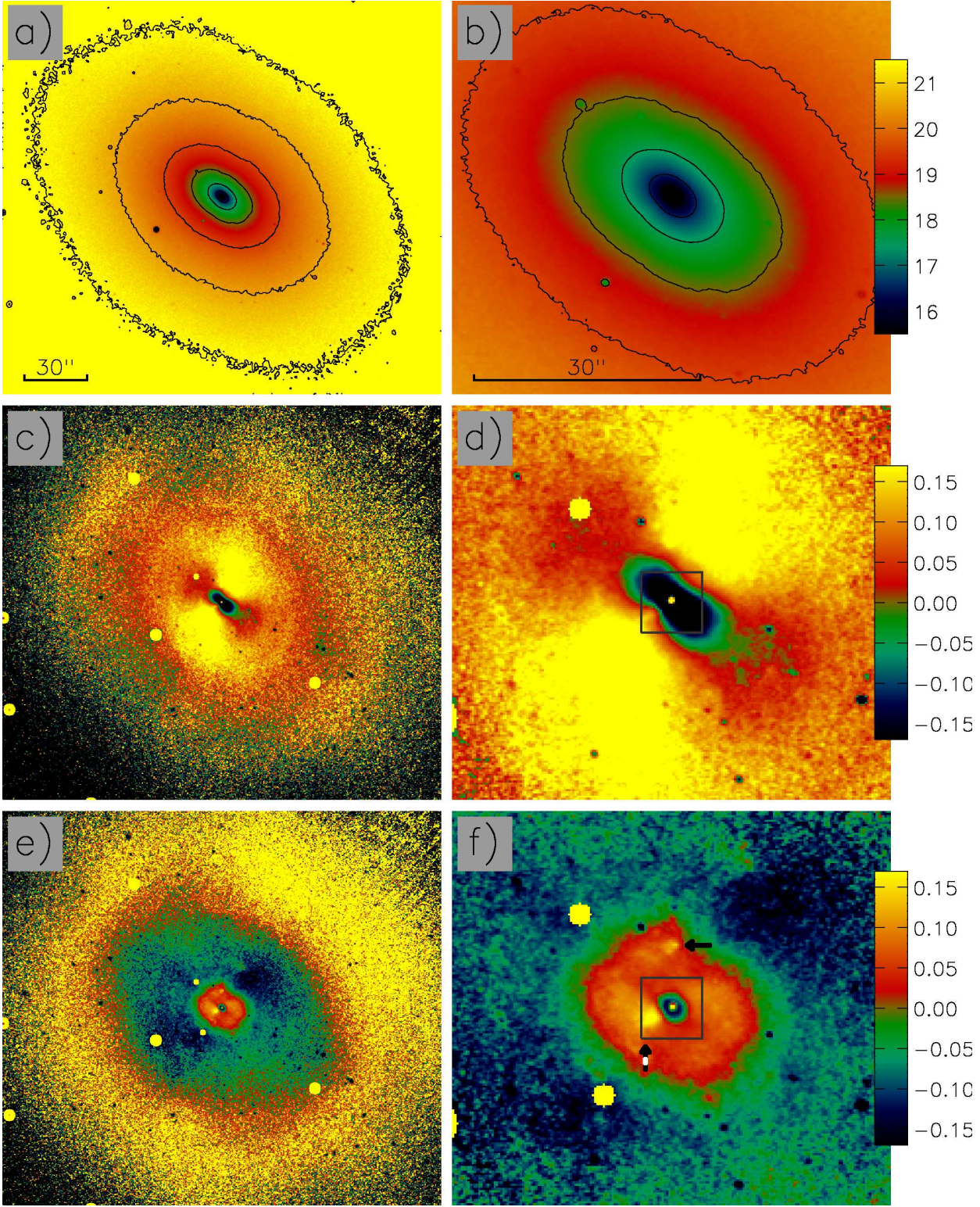}
\caption{SOFI $K\!s$ image of NGC 1316 emphasizing the outer (top left panel, a) and inner (top right panel, b) parts of the galaxy. The middle panels (c,d) show the residual image after subtracting the {\sc budda} bulge+agn model, while the bottom panels (e,f) show the residual image from the bulge+disc+agn model and the spatial scale is the same as in the corresponding top panels. The color coding is indicated in the right-hand bars, in magnitudes per square arcsecond. The square in panels c and f represents the SINFONI field of view. The inner gaseous spiral arms are seen as negative residuals (yellow and red) in panel f. The inner stellar disk is seen as positive residual (blue and green), interior to the gaseous spirals and in the center of the SINFONI field of view. (The very central negative residual is 2 pixels in size, and it is smaller than the seeing FWHM.) The two arrows in panel f show positions in the inner and outer parts of the gaseous spirals where the residuals are strongly negative. They coincide with remarkable features seen in the geometric profiles of Fig. \ref{fig:cuts}, which are consistent with the presence of inner gaseous spiral arms. (North is up, East to the left.)}
\label{fig:modelling}
\end{figure*}

\section{Introduction}

NGC 1316 (Fornax A) is a giant elliptical radio galaxy (Fig. \ref{fig:modelling} - top panels) with pronounced dust
patches, H$\alpha$ filaments, ripples, and loops
\citep{schweizer80,schweizer81}, located in the outskirts of the Fornax cluster,
$3\hbox{$.\!\!^\circ$}7$ away from the central giant elliptical NGC 1399. From
its high luminosity and low central velocity dispersion compared to that
expected from the Faber-Jackson relation \citep{donofrio95}, NGC~1316 has been
considered as a merger remnant \citep{bosma85} with a tidal-tail system
comprising five tails or loops of varying morphology. A merger event seemed to have occurred about 3 Gyr ago
\citep{goud01}, while evidence for two more recent mergers have also been
found \citep{schweizer80,mackfab98}. Cataloged as a S0 peculiar
galaxy, it has been extensively observed in a wide range of wavelengths, and has non-stellar nuclear activity. In the
radio domain, Fornax~A is one of the brightest objects in the sky \citep{ekers83}. It
has prominent giant radio lobes at a position angle of
$\sim$110\hbox{$^\circ$} \citep{wade61} consisting of polarized filaments
\citep{fomalont83} and S-shaped nuclear radio jets \citep{geld84}. Stellar
kinematics of NGC 1316 shows that the galaxy is rotating around the minor axis
at a position angle of $\sim $ $140\hbox{$^\circ$ }$ \citep{bosma85, arna98}.
The kinematics of the outer halo of NGC 1316 indicates that
this early-type galaxy contains as much angular momentum as a giant spiral of
similar luminosity \citep{arna98}.

Decomposition of NGC~1316 radial surface brightness profile has been performed in the past.
The excess of brightness in the
central region (deviations from $r^{1/4}$ law) has been noted in \citet{schweizer80}.
\citet{donofrio01} found a nuclear
disc from ground-based \textit{B}-band observations, while \citet{desouza04}, also using
ground-based optical data, obtained different parameters for this disc.

In this paper, we revisit the central region of NGC 1316. Using ground-based
near-infrared images and kinematics we show that the discovered brightness excess in the central
region of the galaxy (within 2 arcsec) might be explained by the presence of a kinematically
decoupled core.

For the remaining of this paper, we adopt a distance D = 18.6 Mpc, based on HST measurements of
Cepheid variables \citep{madore99} in NGC~1316, so that 1 arcsec
corresponds to 90 pc.

\section{Observations and data reduction}
\label{txt:obs}

The photometric observations were made on the 7th and 8th of March
2001 using the near-IR imager/spectrometer SOFI ($J$,
$H$, and $Ks$ bands) on the 3.5\,m NTT
telescope in La Silla, Chile. The SOFI detector is a Hawaii HgCdTe
$1024\times1024$ array. We used the SOFI large-field mode, with a
pixel scale 0.29 arcsec per pixel and a field of view 4.9 arcmin.
Conditions during the observations were generally clear with
a typical seeing better than 0.7 arcsec.
The integration times were $20\times6$ s for each band.
The observations  were carried out using the ``jitter'' technique.
Jittering was controlled by an automatic jitter template,
which produces a set of dithered frames.
Offsets are generated randomly within a box of
$40\times40$ arsec centered on the galaxy center.
Since the full size of the galaxy observed is bigger than the field of
view of SOFI we obtained intermediate sky frames of the same exposure time.
A classical near-infrared data reduction procedure was adopted.
Dark frame subtraction, flat fielding and illumination correction were
applied using {\sc ccdred} tasks in IRAF\footnote{\emph{Image
Reduction and Analysis Facility} (IRAF) is distributed by NOAO, which
is operated by AURA, Inc., under contract to the NSF.}. The final
sky-subtracted images were aligned and combined using {\sc jitter}
routine of the ESO's {\sc eclipse} package \citep[version 4.9.0,][]{eclipse}.
The zero-points were checked by comparing the
photometry of stars with measurements from 2MASS.

\section{2D image decomposition}
\label{txt:models}

\begin{figure}
\includegraphics[width=0.45\textwidth,clip=true]{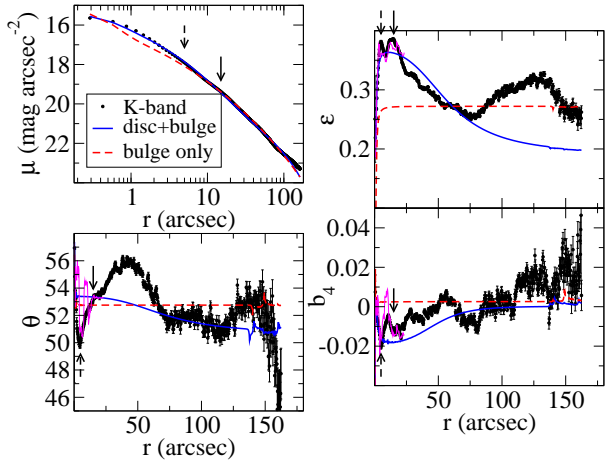}
  \centering
  \caption{Results from ellipse and {\sc budda} fits to the SOFI $K\!s$-band image. The top left panel shows the surface brightness radial profiles
of the galaxy (dots with error bars) and the {\sc budda} models with and without a disc component (blue/solid and red/dashed lines, respectively - both models also include an AGN component). The top right panel is the radial profile of ellipticity, and the corresponding profiles of position angle and
the $b_4$ Fourier coefficient are shown at bottom left and right panels, respectively. The dashed and solid arrows show the position of the features seen in the residual image in Fig. \ref{fig:modelling}f. The thin magenta lines are produced from ellipse fits to an image of the disc+bulge model, in which the model inner spirals arms seen in Fig. \ref{fig:spi} are {\em subtracted} (only the inner part of the ellipse fits is shown -- see text for details). They reproduce well the inner features seen in the galaxy geometric profiles, suggesting the presence of inner gaseous spirals in NGC 1316.}
  \label{fig:cuts}
\end{figure}

\begin{figure}
\includegraphics[width=0.2\textwidth,clip=true]{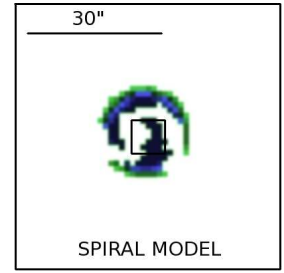}
  \centering
  \caption{Model inner spiral arms built with {\sc iraf} and used to simulate the effects of inner gaseous spiral arms on radial profiles of geometric parameters from ellipse fits to the disc+bulge model of NGC 1316 (see the magenta lines in Fig. \ref{fig:cuts}). The color code is the same as in Fig. \ref{fig:modelling}f. As in Fig. \ref{fig:modelling}f, the inner square shows the SINFONI field of view. (North is up, East to the left.)}
  \label{fig:spi}
\end{figure}

\begin{figure*}
\centerline{
\includegraphics[width=6cm]{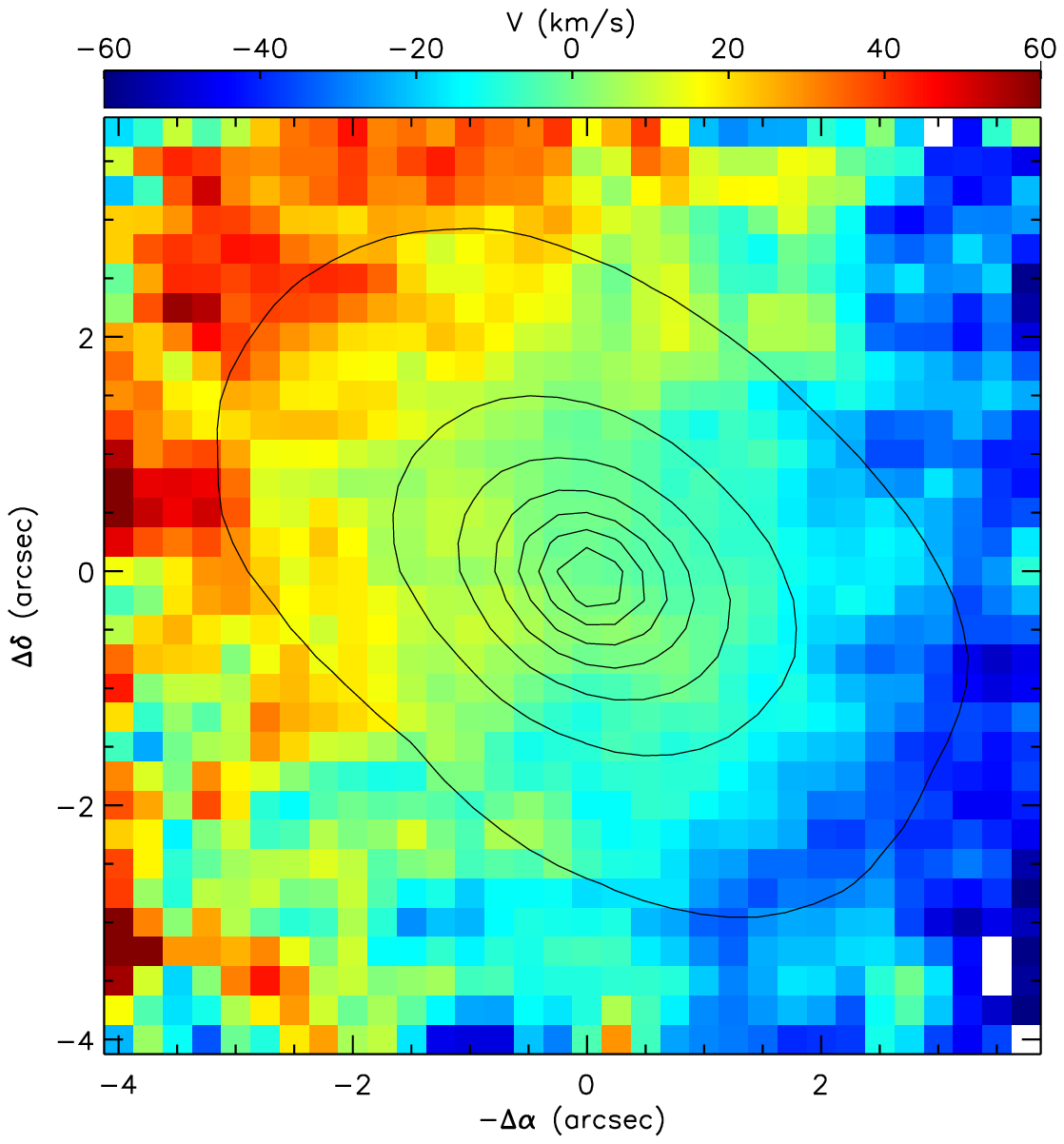}
\includegraphics[width=6cm]{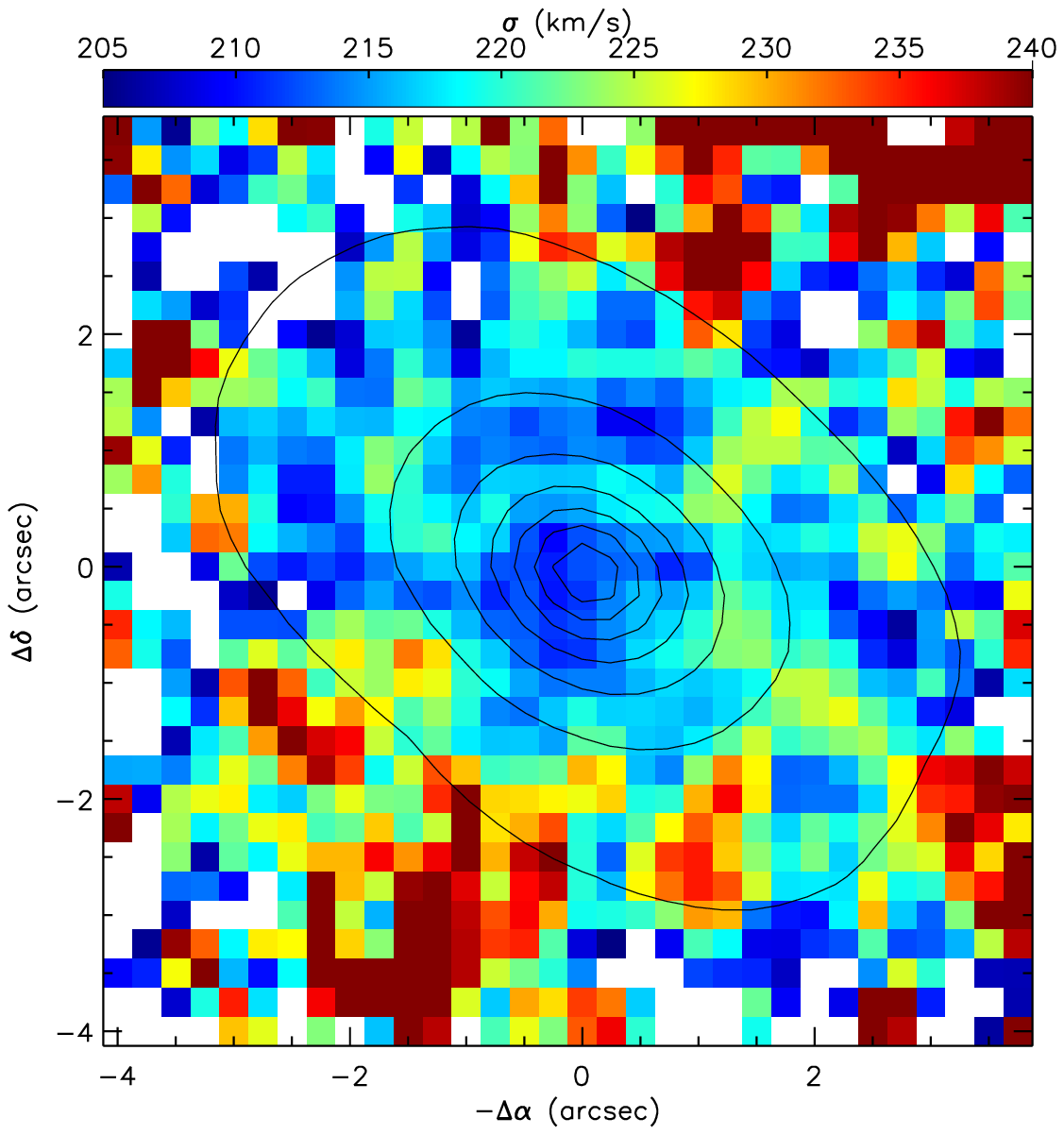}
\includegraphics[width=6cm]{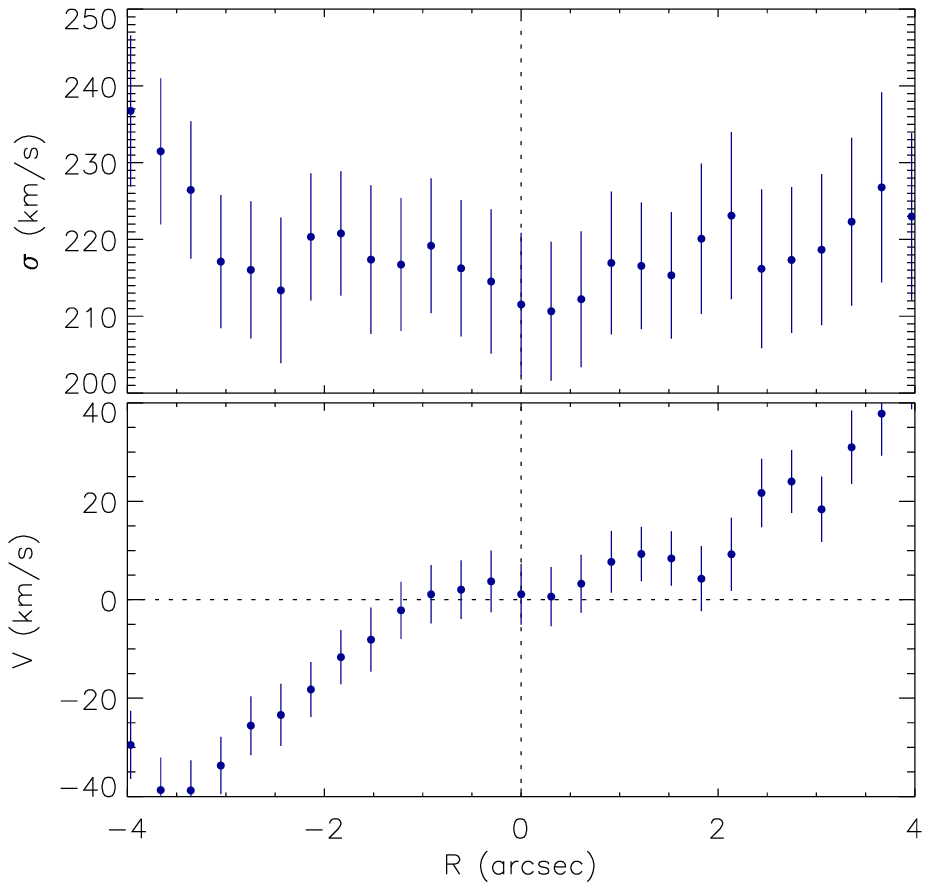}
}
  \caption{The SINFONI kinematic data. Velocity field of stars (left) and line-of-sight velocity dispersion map (middle) with continuum near $2.1\mu m$ contours superimposed. The right panel shows the cuts through  the velocity and velocity dispersion maps along the major axis ($PA=53^\circ$). The systemic velocity is  1785\,km~s$^{-1}$.}
  \label{fig:sinfoni}
\end{figure*}

We used {\sc budda} \citep{desouza04,gad08} to produce 2D image models of NGC 1316, using the SOFI $K\!s$-band image. The PSF is modeled using several suitable stars in the field and using a Moffat function, which is known to produce better results than a Gaussian \citep{tru01}. The first model corresponds to a single component, described with a S\'ersic surface brightness radial profile and fixed position angle and ellipticity, plus a point source, convolved with the PSF, to model the AGN. The resulting fit indicates a S\'ersic index $n=4.2$ and an effective radius of 96 arcsec. Since the inner half of the galaxy seems somewhat boxy, we left the boxiness parameter $c$ free, but that resulted in $c=2$, corresponding to a perfect ellipse (i.e. neither boxy or disky). We obtained a residual image by subtracting this model from the original galaxy image, and verified that a significant inner structure is seen (middle panels of Fig. \ref{fig:modelling}). We produced test fits with an offset centroid to check whether this feature could be a result of an ill-defined centroid, but we find that the uncertainty in the centroid position is of 1 pixel only. In any case, a bad centroid position would produce a region of negative residual alongside another of positive residual, a pattern which is different from what is seen in Fig. \ref{fig:modelling}d. This panel shows an oscillation between negative and positive residuals. This also cannot be the product of a mismatch between the PSF in the image and how we modeled the PSF, as the residuals are an order of magnitude larger than the PSF. This motivated us to produce a second model, which includes a disc component following an exponential surface brightness radial profile. The best solution model in this case has a disc component with a scale length of 58 arcsec, and a bulge component with S\'ersic index $n=2.9$ and effective radius of 17 arcsec. The bulge component is now slightly boxy, with $c=2.2$, and the bulge/total ratio is 0.37. This suggests that by adding another component the code is able to retrieve better the boxy aspect of the bulge. This model produces a much better fit to the galaxy image, as can be seen in Fig. \ref{fig:cuts}, which shows results from ellipse fits to the galaxy image and the models, using {\sc iraf}. The reduced $\chi^2$ goes down from 5.5 to 3.8 with a total of 12 parameters fitted, instead of 8 (4 new parameters to describe the disc). Further, the decomposition with only the bulge model results in a residual image with positive residuals amounting to $\sim 16\%$ of the total galaxy light.

The residual image from the first {\sc budda} model (Fig. \ref{fig:modelling}d) and the ellipse fits (Fig. \ref{fig:cuts}) give a strong argument for the presence of an extra inner component in NGC 1316. The difficulty in interpreting such substructure is that it does not resemble morphologically a well-know stellar structure. This might be a result of the inadequate fit to the galaxy image using only one component.
An extra component in any model usually leads to a better fit, even if such component in the model does not have a straightforward correspondence to the real structural component in the galaxy. In other words, the disc in the second {\sc budda} model might be truly fitting just an extra component (or components) which is not necessarily a usual disc as in disc galaxies. Thus, a conservative interpretation is that the extra disc component in the second model corresponds to substructures which result in the morphologically perturbed appearance of NGC 1316 (which are well described by an exponential profile), and not to a proper stellar disk. This in turn is consistent with a recent merger event.

Interestingly, the residual image from the second {\sc budda} model (Fig. \ref{fig:modelling}f) reveals a negative residual (i.e. a region in which the model is brighter than the galaxy) of less than about 0.2 mag, which resembles inner spiral arms. It is tempting to interpret this feature as inner {\em gaseous} spirals, which could be produced by the accretion of a companion galaxy. To verify this interpretation, we produced a new synthetic image as follows. We created a model of spiral arms resembling those seen in the negative residual of Fig. \ref{fig:modelling}f, and {\em subtract} it from the {\sc budda} model image in our second fit. These model inner spiral arms are shown in Fig. \ref{fig:spi}. They were built using {\sc artdata} tasks in {\sc iraf}, manually edited with the task {\sc imedit}, and convolved with a Gaussian, with the same FWHM as the PSF in our SOFI $Ks$-band images, using the task {\sc gauss}. Their surface brightness levels have Poisson fluctuations added, decline roughly exponentially from the centre, and are chosen as to mimic the residual levels seen in Fig. \ref{fig:modelling}f, i.e. below 0.2 mag per square arcsecond. We can thus mimic the effects of gaseous spirals attenuating the light emission from the central region of NGC 1316. We produced ellipse fits to this new synthetic image, and the results are also shown in Fig. \ref{fig:cuts} (magenta lines). The ellipse fits to our new synthetic image reproduce well the oscillations in ellipticity, position angle and $b_4$ seen in the inner part of NGC 1316. This supports our interpretation, suggesting the presence of inner gaseous spiral arms in the galaxy.
It is important to note that interior to these spirals, there is a region of {\em positive} residuals of about 2", which can be an inner {\em stellar} disc. This will be discussed further below.

A similar photometric analysis has been recently done by \citet{lanz10} using Spitzer archival data. They performed a 2D fit of the 3.6 micron image, using a bulge-only model, and find very similar residuals, as compared to our first fit (see their figure 1c). However, they did not produce a fit with a disc component. They find structural parameters somewhat different than those from our first fit. They find a S\'ersic index of $6.1\pm0.1$, while we find a value of $4.2\pm0.9$. They find an effective radius of $146\pm1$ arcsec, whereas we find $96\pm7$ arcsec. It can be that a deeper Spitzer image results in a lager effective radius. Nevertheless, a profile with S\'ersic index 6 does not differ terribly from one with an index of 4 \citep[see figure 1 in][]{graham05}. They do not show how their modeled profile matches the observed one. The residuals we find in Fig. \ref{fig:modelling}f only appear when doing the second fit, including the disc. We note that the pixel scale and the PSF FWHM are significantly better in our SOFI images, as compared to the Spitzer 3.6 micron images. Our pixel scale is 0.29 arcsec/pix and our PSF FHWM is 0.7". The Spitzer image has 1.2 arcsec/pix and a PSF FWHM of 1.7".

\section{SINFONI kinematics}

Information on stellar dynamics is crucial to understand the origin of the {\em stellar} disc-like structure suggested to be present, in the inner 2", in the previous section. If the observed NIR brightness excess is really associated with the excess of stellar density we can expect that this region has also decoupled features in NIR kinematics.

The most detailed study of the NGC~1316 stellar kinematics, based on VLT/SINFONI integral-field data, was recently published by \citet{nowak08}. Their investigation was focused on the very center of the galaxy, where  gravitational forces of the supermassive black hole are important. Nina Nowak kindly provided us with reduced SINFONI spectra taken with the largest field-of-view (mentioned as ``250 mas data'' in their original paper). The spectral data were presented in a data cube, where each 0.25 arcsec spaxel in the $8\times8$ arcsec field has a galaxy spectrum in the range $1.95-2.45\,\mu m$.
The stellar kinematics information were extracted   using IDL-based software and algorithms developed to study stellar kinematics  with the integral-field spectrograph MPFS \citep{moiseev2004}. The velocity field and velocity dispersion map were created by cross-correlation with the spectrum of a K5/M0III template star (HD181109) in the wavelength range $2.260-2.343\mu m$, which contains high contrast $^{12}$CO absorption bands. Unlike \citet{nowak08}, we did not use a radial or azimuthal binning of the original data, in order to keep the original spatial sampling in the data cube.

The maps of line-of-sight velocities ($v$) and velocity dispersion ($\sigma$) of stars derived from the SINFONI data  are displayed in Fig.~\ref{fig:sinfoni}. These maps reveal two kinematic features that are  different from the simple picture expected for a rotating oblate spheroid. The first one is an absence of clear rotation in the central region ($r<1.5-2$ arcsec). It seems as a central ``plateau'' of a constant color in the velocity field. The velocity field cross-section along the galaxy major axis (see right panel in Fig.~\ref{fig:sinfoni}) shows that the circumnuclear velocity distribution is complex -- a hint of low-amplitude counter-rotation  appears on $r<1$ arcsec. The second surprising kinematic feature is a slight central depression of the velocity dispersion on $r<2$ arcsec. This feature has a low contrast, but it is visible as a central ``blue spot'' on the velocity dispersion map and also as a minimum in the $\sigma$ radial distribution along the major axis. \citet{nowak08} discussed in detail the possible origin of the ``$\sigma$-drop'' feature using their high spatial resolution ``25 mas'' SINFONI data set. They have considered different effects (data reduction artifacts, stellar population, AGN emission, etc.) and  concluded that the $\sigma$-drop observed  in the high resolution data  could be caused by  ``either the AGN continuum emission or a cold stellar subsystem or both''. However, they showed that an AGN distorts the stellar kinematics only in very center of the galaxy ($r\leq0.06$ arcsec). Therefore, we suggest that the $\sigma$-drop on a larger radial scale is caused by a dynamically cold stellar population. We note that both kinematic features (an absence of central rotation and a $\sigma$-drop) were already mentioned by \citet{nowak08} and our kinematic maps show a good agreement with their figure 6. However, these features are more clearly visible in our un-binned maps. We emphasize that the spatial scale of the decoupled features in the stellar kinematics agrees very well with the value expected from the NIR brightness excess found above ($r\leq2$ arcsec). Therefore, the connection between circumnuclear morphological and kinematical features is evident.

\section{Discussion and conclusions}

A multicomponent structure in the central regions of early-type galaxies is
a well-known phenomenon  \citep{Rest01,ES03}. Moreover, the presence of dust in the vicinity of central regions of post-merger galaxies (NGC 1316 is one of the best examples) makes a study of those regions even more complicated.

A detailed analysis of the NIR structure of NGC 1316 (Figs. \ref{fig:modelling} and \ref{fig:cuts}) shows a negative residual suggestive of inner {\em gaseous} spiral arms from a radius of about 5" to 15". Numerical simulations of galaxy mergers, such as those by \cite{dimatteo07}, often show the formation of a gaseous disc with gaseous spiral arms and bars. We note a similar structure in the center of other giant elliptical radio galaxy, NGC 5128 (Centaurus A), observed in the mid-infrared by \citet{mirabel99}.

We thus put forward the interpretation that the structure we find near the center of NGC 1316 is indeed a gaseous component, most likely formed from in-falling material from a recent accretion event, which attenuates the background stellar light from the galaxy. We note that \citet{horellou01} detected CO, presumably associated to dust, near the center of the galaxy, although at a somewhat larger spatial scale than that of the structure we are investigating here. Their CO map shows a structure which does not seem correlated with the negative residuals we find, but such a comparison is not straightforward, as their spatial resolution (22'') is rather coarse. \citet{horellou01} also advocate that the CO they found resulted from a recent accretion. It is worth noting that \citet{nowak08} found a weak signature of molecular Hydrogen in a region northeast of the nucleus, a result which is corroborated by our own analysis of the SINFONI spectra. Another plausible interpretation is that the structure we see is caused by the galaxy AGN jet. Although we cannot rule out this possibility, we note that no emission lines typical of AGN are found in the SINFONI spectra, which is in principle inconsistent with it.

Interior to these spirals ($\leq$ 2"), we find another structure, resembling a nuclear {\em stellar} disc (inside the square representing the SINFONI field of view in Fig. \ref{fig:modelling}f).
The stellar kinematics (Fig. \ref{fig:sinfoni}) suggests the presence of a kinematically decoupled core on the same spatial scale. This suggests that indeed the nuclear stellar structure we find in the structural analysis is a kinematically cold stellar disc.

This nuclear disc could also be formed via a recent merging event. \cite{donofrio95} noted that the central velocity dispersion in NGC 1316 is lower than the one expected for a galaxy of this luminosity from the Faber--Jackson relation (260 versus 400 km/s). It could be possibly explained by the presence of a dynamically cold disc-like component. Further evidence is presented by \citet[][their figure 8]{nowak08}, which found indications of a younger stellar component in the very center ($\leq$ 0.5") of NGC 1316.

\section*{Acknowledgments}

AM would like to thank the Russian Foundation for Basic Research (project  09-02-00870). We are also grateful to Ralf Bender and Olga Sil'chenko for useful comments, and an anonymous referee for important remarks that improved the paper.
This publication makes use of data products from the Two Micron All Sky
Survey, which is a joint project of the University of Massachusetts
and the Infrared Processing and Analysis Center/California Institute
of Technology, funded by the National Aeronautics and Space
Administration and the National Science Foundation. This research has made use of the NASA/IPAC Extragalactic Data base (NED) operated by the Jet Propulsion Laboratory, California Institute of Technology, under contract with the National Aeronautics and Space Administration. The research is partly based on data obtained from the Multimission Archive at the Space Telescope Science Institute (MAST). STScI is operated by the Association of Universities for Research in Astronomy, Inc., under NASA contract NAS5-26555.

\bsp

\label{lastpage}

\end{document}